\documentstyle[aps,epsfig,floats,preprint,amsfonts]{revtex}
\begin{document}
\draft 
\preprint{\vbox{{\hbox{\tt SOGANG-HEP 303/02 gr-qc/0301004}}}}
\title{Duality of Quasilocal Gravitational Energy and Charges with Non-orthogonal Boundaries}
\author
{Sung-Won Kim$^{a}$\footnote{electronic address:sungwon@mm.ewha.ac.kr},
Won Tae Kim$^{b}$\footnote{electronic address:wtkim@ccs.sogang.ac.kr}, John
  J. Oh$^{b}$\footnote{electronic address:john5@string.sogang.ac.kr}, and
  Ki Hyuk Yee$^{b}$\footnote{electronic address:quicksilver@string.sogang.ac.kr}}
\address{${}^{a}$ Department of Science Education, Ewha Women's University, Seoul 120-750, Korea\\
         ${}^{b}$ Department of Physics and Basic Science Research Institute, Sogang University, C.P.O. Box 1142, Seoul 100-611, Korea}
\maketitle
\begin{abstract}
We study the duality of quasilocal energy and charges with
non-orthogonal boundaries in the (2+1)-dimensional low-energy string
theory. Quasilocal quantities shown in the previous work and some
new variables arisen from considering the non-orthogonal boundaries as
well are presented, and the boost relations between those quantities
are discussed. Moreover, we show that the dual properties of
quasilocal variables such as quasilocal energy density, momentum
densities, surface stress densities, dilaton pressure densities, and
Neuve-Schwarz(NS) charge density, are still valid in the moving
observer's frame. 
\end{abstract}
\pacs{PACS : 04.20.-q, 04.90.+e}
\bigskip
\newpage

\section{Introduction\hfil{}}
The study of a gravitational system with finite boundaries gives some
advantages rather than that with the asymptotic fall-off behavior such
as asymptotic flatness. First, generically, treating a gravitational system with a bounded and finite spatial
region should be independent of the asymptotic behavior of the gravitational
field. Therefore, this kind of study is considerably useful for
developing a theoretical formulation which is irrelevant to the
specific asymptotic properties of the system such as an asymptotic
flatness. Second, if one constructs a gravitational partition function
without any inconsistencies by assuming finite boundaries, then the
construction of the gravitational partition function is only possible
when the system with a finite size is stable. For example, the heat
capacity for the Schwarzschild black hole is negative if the temperature
at asymptotic region is fixed and the partition function for the
black hole is not consistent. However, if we consider the fixed
temperature at a finite spatial boundary, the heat capacity is
positive and the partition function is well-defined. Third, from the
physical viewpoint, one can define thermodynamics which is appropriate to
observers placed at a finite region from black holes. In these
respects, it is meaningful to define thermodynamic quantities
appropriately at a finite boundary.

Some years ago, Brown and York have studied the quasilocal quantities
such as the quasilocal energy, angular momentum, and spatial
stress through the Hamilton-Jacobi analysis of a gravitational system
\cite{by}. Those quantities are closely related to the first law of
black hole thermodynamics through the path integral formulation of
gravitational system \cite{by2}. This formalism was extended to
include the most general case of gauge fields coupled to the dilaton
gravity in the context of string theories \cite{cm}, and
the temperature, energy, and heat capacity of AdS black holes have been
studied by use of this formulation in Ref. \cite{bcm}. The Hamiltonian
and entropy in asymptotically flat spacetimes(AFS) and anti-de
Sitter(AdS) have been studied in Ref. \cite{ht}, and the relevant issues for the two-dimensional black hole \cite{cm2} and the
quasilocal thermodynamics of Kerr-AdS(K-AdS) and Kerr-de Sitter(K-dS)
\cite{deh} were also intensively investigated.

However, Brown and York's quasilocal formulation is based on the
assumption that the spacetime foliation is orthogonal to the timelike
boundary, which describes the quasilocal quantities seen by static
observers in a weak gravitational field, and it seems to be a somewhat strong
restriction. When one takes into
account finite spatial boundaries in a strong gravitational field,
gravitational force acts on each spatial boundary with a different extent.
Therefore, in general, the unit normal defined on the hypersurface at
a certain time is not orthogonal to the unit normal defined on the
finite spatial boundary, and it is too difficult to
calculate the quasilocal quantities seen by observers who are falling into a
black hole through the quasilocal formulation with orthogonal boundaries.
To generalize the formulation and overcome this difficulty,
Booth and Mann reformulated the quasilocal analysis in the presence of
non-orthogonal boundaries \cite{bm}, and the related works appear in Ref. \cite{nonorth} 

On the other hand, in the context of string theory, duality is
considered as a symmetry which relates a certain solution to another
one. In the (2+1)-dimensional low-energy string theory, this duality is more meaningful
in that the dual solution of the  Ba$\tilde{{\rm
    n}}$ados-Teitelboim-Zanelli(BTZ)\cite{btz} black hole is known as
the (2+1)-dimensional charged black string \cite{hw}. And the duality of
the quasilocal quantities between these dual solutions and
quasilocal thermodynamics of the dilatonic gravitational system with orthogonal
boundaries was studied in Ref. \cite{hkp}. The quasilocal energy density and
its dual are invariant under the dual transformation while the quasilocal angular momentum density and its dual are
interchanged with the quasilocal Neuve-Schwarz(NS) charge and its
dual. In addition, the dual invariance between the surface spatial
stress density and the dilaton pressure density appears in the
combination of both quantities as ${\cal E} = {\cal E}^{d}$, ${\cal J}_{\phi} = - ({\cal Q}^{d})^{\phi}$, $({\cal Q})^{\phi} = -{\cal J}^{d}_{\phi}$, $
{\cal S}^{ab} \delta\sigma_{ab} + \Upsilon \delta\Phi = {\cal
  S}_{d}^{ab} \delta\sigma_{ab}^{d} + \Upsilon_{d} \delta\Phi^{d}$,
where ${\cal E}$, ${\cal J}$, ${\cal Q}$, ${\cal S}^{ab}$, $\cal
Y$, $\sigma_{ab}$, and $\Phi$ are the quasilocal surface energy density, the quasilocal momentum
density, the quasilocal NS charge density, the quasilocal spatial
stress density, the quasilocal pressure density, the surface spatial stress
tensor, and the dilaton field, respectively. 

In this paper, we shall study the dual properties of quasilocal quantities
for the (2+1)-dimensional dilatonic gravity with
non-orthogonal boundaries. In Sec. \ref{sec:II}, the notations and the
setup
for the double-foliation of quasilocal formalism with non-orthogonal
boundaries are presented. The unit vectors
normal to both spatial and temporal boundaries are defined and
splittings of extrinsic curvatures on the spacelike hypersurface and
spatial boundary are obtained by the definitions of the induced
metrics and the extrinsic curvatures. 
The quasilocal variables with
non-orthogonal boundaries and their boost relations, and dual
properties between those variables are given in Sec. \ref{sec:III}. In
Sec. \ref{sec:IV}, some concluding remarks and discussions on our results follow.

\section{Preliminary : Notations and Setup \hfil{}}\label{sec:II}
In this section, we present a double-foliation for
Arnowitt-Deser-Misner(ADM) splitting of the
metric and the corresponding some kinematics. Then we shall discuss
the notations and extrinsic curvature splittings for quasilocal
formalism with non-orthogonal boundaries.

Generically, when we take into account a finite spatial boundary on
manifold ${\cal M}$ in a strong gravitational field such as an
adjacent region of black hole horizon, each boundary is exposed to a
different gravitational force. This fact enhances the motivation of
the generalized quasilocal formalism, which can be possible by
considering non-orthogonal boundaries.

\begin{figure}[htbp]
    \begin{center}
    \leavevmode
    \centerline{
        \epsfig{figure=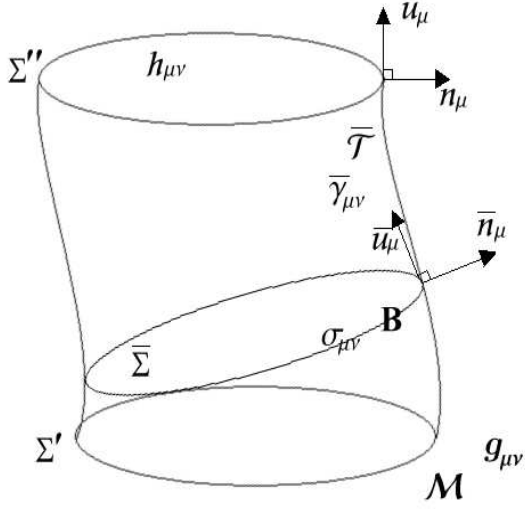, width=8cm, height=8.5cm}
        }
    \caption{Spacetimes Foliation : The spacetime manifold ${\cal M}$
    which is topologically $\Sigma\times\bar{\cal T}$ can be
    foliated by spatial and temporal boundaries denoted by $\bar{\cal
    T}$ and $\Sigma$, respectively. On each boundary, the unit
    normal vector, induced metric, and extrinsic curvature are defined.} 
    \label{foliation}
    \end{center}
\end{figure}

Let us consider a double-foliation of spacetime manifold ${\cal M}$ with
spatial and temporal boundaries as shown in FIG. \ref{foliation}. Then
we can take $t=const$ and $s=const$ surfaces on boundaries
$\Sigma$ and $\bar{\cal T}$, and the unit normal vectors are defined as
$u_{\mu}=-N\nabla_{\mu}t$ on $\Sigma$ and
$\bar{n}_{\mu}=\bar{M}\nabla_{\mu}s$ on $\bar{\cal T}$, where $N$ and
$\bar{M}$ are normalization functions determined by satisfying $u\cdot
u=-1$ and ${\bar{n}}\cdot{\bar{n}} = 1$. On boundaries
of $\Sigma$ and $\bar{\cal T}$, the induced metrics $h_{\mu\nu}$, $\bar{\gamma}_{\mu\nu}$ and the
corresponding extrinsic curvatures $K_{\mu\nu}$,
$\bar{\Theta}_{\mu\nu}$ can be defined as
\begin{eqnarray}
  \label{indmetrics}
  h_{\mu\nu} &=& g_{\mu\nu} + u_{\mu}u_{\nu} ~~(\rm on~ \Sigma),\\
  \bar{\gamma}_{\mu\nu} &=& g_{\mu\nu} -
  \bar{n}_{\mu}\bar{n}_{\nu}~~(\rm on~ \bar{\cal T}),
\end{eqnarray}
and
\begin{eqnarray}
 \label{extcurv}
 K_{\mu\nu} &=& - h_{\mu}^{\alpha} \nabla_{\alpha} u_{\nu} ~~(\rm on~
 \Sigma), \\
 \bar{\Theta}_{\mu\nu} &=& - \bar{\gamma}_{\mu}^{\alpha}
 \nabla_{\alpha} \bar{n}_{\nu} ~~(\rm on~\bar{\cal T}).
\end{eqnarray}
And we can define new unit vectors $n_{\mu}$ and $\bar{u}_{\mu}$ as
$n_{\mu}=MD_{\mu}s = \gamma^{-1}h_{\mu}^{\nu}\bar{n}_{\nu}$ and
$\bar{u}_{\mu}=-\bar{N}{\cal D}_{\mu}t =
\gamma^{-1}\bar{\gamma}_{\mu}^{\nu}u_{\nu}$, where $D_{\mu}$ and
${\cal D}_{\mu}$ are covariant derivatives projected into $\Sigma$ and
$\bar{\cal T}$ surfaces, and the boost factor
$\gamma=(1-v^2)^{-1/2} = M/\bar{M} = N/\bar{N}$, where $v$ is a proper
radial velocity. From these relations, the relations between unit
normal vectors seen by ``barred'' frame and ``unbarred frame'' are obtained as
\begin{eqnarray}
  \label{relunit}
  \bar{u}_{\mu} = \gamma u_{\mu} + \gamma v n_{\mu}, \nonumber\\
  \bar{n}_{\mu} = \gamma n_{\mu} + \gamma v u_{\mu}.
\end{eqnarray}

On the boundary $B$, the induced metric is given in two ways as
$\sigma_{\mu\nu}=g_{\mu\nu} + u_{\mu}u_{\nu} - n_{\mu}n_{\nu} =
g_{\mu\nu} + \bar{u}_{\mu}\bar{u}_{\nu} - \bar{n}_{\mu}\bar{n}_{\nu}$
and the extrinsic curvatures are also defined as $k_{\mu\nu} =
-\sigma_{\mu}^{\alpha}\sigma_{\nu}^{\beta}\nabla_{\alpha}n_{\beta}$
and $\ell_{\mu\nu} = -
\sigma_{\mu}^{\alpha}\sigma_{\nu}^{\beta}\nabla_{\alpha}u_{\beta}$.
Note that the notations used in this paper for the foliation of spacetimes
are summarized in TABLE. \ref{table1}.

On the other hand, the extrinsic curvature on $\bar{\cal T}$ boundary,
$\bar{\Theta}_{\mu\nu}$, can be splitted by the extrinsic curvatures
on $B$ boundary, $k_{\mu\nu}$ and $\ell_{\mu\nu}$ as,
\begin{equation}
  \label{splitting1}
  \bar{\Theta}_{\mu\nu} = \gamma k_{\mu\nu} + \gamma v \ell_{\mu\nu} +
  (\bar{n}\cdot \bar{a})\bar{u}_{\mu}\bar{u}_{\nu} + 2
  \sigma_{({\mu}}^{\alpha} \bar{u}_{{\nu})}(n^{\lambda}
  K_{\alpha\lambda} - \gamma^2 \nabla_{\alpha} v),
\end{equation}
where $\bar{a}_{\mu}=\bar{u}^{\alpha}\nabla_{\alpha}\bar{u}^{\mu}$ is
the acceleration of $\bar{u}^{\mu}$. Similarly, the splitting of the extrinsic
curvature on the ${\Sigma}$ boundary $K_{\mu\nu}$, is obtained as
\begin{equation}
  \label{splitting2}
  K_{\mu\nu} = \ell_{\mu\nu} + (u\cdot b)n_{\mu}n_{\nu} +
  2\sigma_{({\mu}}^{\alpha}n_{{\nu})}K_{\alpha\lambda}n^{\lambda},
\end{equation}
by use of the extrinsic curvature on $B$ boundary $\ell_{\mu\nu}$ and
the acceleration $b^{\mu} = n^{\alpha} \nabla_{\alpha} n^{\mu}$ of $n^{\mu}$.
\begin{table}
\caption{Notations for foliation of spacetimes ${\cal M}$}
\begin{tabular}{lcccccc}\label{table1}
  contents & metric & \vbox{\hbox{covariant}\hbox{derivative}} & \vbox{\hbox{unit}\hbox{normal}} & \vbox{\hbox{intrinsic}\hbox{curvature}} & \vbox{\hbox{extrinsic}\hbox{curvature}} & momentum  \\ \tableline
spacetimes $\cal{M}$  & $g_{\mu\nu}$ & $\nabla_{\mu}$ & &
  $R_{\mu\nu\kappa\lambda}$ & &  \\
spacelike hypersurface $\Sigma$ & $h_{ij}$ & $D_{i}$ & $u_{\mu}$ &
  ${\cal{R}}_{ijkl}$ & $K_{ij}$ & $P^{ij}$ \\ 
timelike hypersurface $\bar{\cal{T}}$  & $\bar{\gamma}_{ij}$ & ${\cal{D}}_{i}$
  & $\bar{n}_{\mu}$ & & $\bar{\Theta}_{ij}$ & $\bar{\Pi}_{ij}$\\
boundary $B=\Sigma\cap\bar{\cal T}$ & $\sigma_{ab}$ & & & & $k_{ab}$, $\ell_{ab}$ & \\
\end{tabular}
\end{table}

\section{Duality of Quasilocal Quantities with Orthogonal and
  Non-orthogonal Boundaries \hfil{}}\label{sec:III}
\subsection{Static observers and duality of quasilocal
  quantities}
The dilatonic action coupled with NS-NS
field strength in (2+1)-dimensions is given as
\begin{eqnarray}
  \label{action}
  S &=& \frac{1}{2\pi} \int_{{\cal M}} d^3x \sqrt{-g}\Phi\left[ R +
  \Phi^{-2} (\nabla \Phi)^2 + \frac{4}{l^2} -
  \frac{1}{12}H^2\right]  \nonumber\\
    &+& \frac{1}{\pi} \int_{\Sigma} d^2x\sqrt{h}\Phi K -
  \frac{1}{\pi}\int_{\cal T} d^2x \sqrt{-\gamma} \Phi \Theta,
\end{eqnarray}
where $-1/2\ln\Phi$ is a dilaton field, $H$ is a three-form field strength
of the anti-symmetric two-form field $B$ with $H=dB$, and $l^{-2} =
-\Lambda$ is a negative cosmological constant.

The variation of action (\ref{action}),
\begin{eqnarray}
  \label{variation}
  \delta S &=& \int_{\cal M} d^3x \sqrt{-g}\left[
  \left(\Xi_{G}\right)_{\mu\nu} \delta g^{\mu\nu} + \left(\Xi_{\rm dil}
  \right) \delta \Phi + \left(\Xi_{\rm NS} \right)^{\mu\nu} \delta
  B_{\mu\nu}\right] \nonumber \\
  &+& \int_{\Sigma}d^2x \left[ P^{ij}\delta h_{ij} + P_{\rm dil} \delta \Phi +
  P_{\rm NS}^{ij} \delta B_{ij}\right] \nonumber \\
  &+& \int_{\cal T}d^2x \left[\Pi^{ij} \delta\gamma_{ij} +
  \Pi_{\rm dil} \delta \Phi + \Pi_{\rm NS}^{ij} \delta B_{ij}\right],
\end{eqnarray}
gives the equations of motion,
\begin{eqnarray}
  \label{eqnmot}
  & &2\pi (\Xi_{G})_{\mu\nu} = \Phi G_{\mu\nu} +
  \nabla_{\mu}\nabla_{\nu} \Phi - g_{\mu\nu} \Box\Phi -
  \frac{1}{2}g_{\mu\nu}\Phi^{-1}(\nabla\Phi)^2 -
  \frac{2}{l^2}g_{\mu\nu}\Phi \nonumber \\
  & &\qquad\qquad - \frac{1}{24} g_{\mu\nu}\Phi H^2 +
  \frac{1}{4}\Phi H_{\mu\lambda\sigma}H_{\nu}^{\lambda\sigma}\nonumber,
  \\
  & &2\pi (\Xi_{\rm dil}) = R + \Phi^{-2}(\nabla\Phi)^2
  -2\Phi^{-1}\Box\Phi + \frac{4}{l^2} - \frac{1}{12}H^2\nonumber, \\
  & &4\pi(\Xi_{\rm NS})^{\mu\nu} = \nabla_{\lambda}(\Phi
  H^{\mu\nu\lambda}),
\end{eqnarray}
where $G_{\mu\nu} = R_{\mu\nu} - 1/2g_{\mu\nu} R$ is the Einstein
tensor. The conjugate momenta on $\Sigma$ and ${\cal T}$
boundaries are given as
\begin{eqnarray}
  \label{momS}
  & &P^{ij} = -\frac{\sqrt{h}}{2\pi} \left[ \Phi(K^{ij} - h^{ij}K) +
  h^{ij}u^{\alpha}\nabla_{\alpha}\Phi\right],\nonumber \\
  & &P_{\rm dil} = - \frac{\sqrt{h}}{\pi} \left[ \Phi^{-1}
  u^{\alpha}\nabla_{\alpha}\Phi - K \right], \nonumber \\
  & &P_{\rm NS}^{ij} = \frac{\sqrt{h}}{4\pi} \Phi u^{\alpha}H^{ij}_{\alpha}
\end{eqnarray}
and
\begin{eqnarray}
  \label{momT}
  & &\Pi^{ij} = \frac{\sqrt{-\gamma}}{2\pi}\left[\Phi(\Theta^{ij} -
  \gamma^{ij}\Theta) + \gamma^{ij}
  n^{\alpha}\nabla_{\alpha}\Phi\right],\nonumber \\
  & &\Pi_{\rm dil} =
  \frac{\sqrt{-\gamma}}{\pi}(\Phi^{-1}n^{\alpha}\nabla_{\alpha}\Phi -
  \Theta)\nonumber, \\
  & &\Pi_{\rm NS}^{ij} = -\frac{\sqrt{-\gamma}}{4\pi}\Phi n^{\alpha}H^{ij}_{\alpha},
\end{eqnarray}
respectively. Especially, the momenta on ${\cal T}$ boundary are
closely related to the quasilocal quantities within this boundary. To
specify these quantities, it is useful to decompose the induced metric
$\gamma_{ij}$ into some projections normal and onto foliation as follows,
\begin{equation}
  \label{ADMdecom}
  \delta \gamma_{ij} = -\frac{2}{N} u_{i}u_{j} \delta N - \frac{2}{N}
  u_{({i}}\sigma_{{j})a} \delta V^{a} +
  \sigma_{({i}}^{a}\sigma_{{j})}^{b} \delta \sigma_{ab},
\end{equation}
where $N$ is a lapse function and $V^{a}$ is a shift vector on ${\cal
  T}$ boundary. As for the NS field $B_{ij}$, this potential on the
  boundary ${\cal T}$ can be written as $B_{ij} = 2
  u_{[i}\sigma^{a}_{{j}]}C_{a} +
  \sigma^{a}_{[{i}}\sigma^{b}_{{j}]}D_{ab}$, where $C_{a} =
  \sigma^{i}_{a} B_{ij} u^{j}$ and $D_{ab} = \sigma^{i}_{a}
  \sigma^{j}_{b}B_{ij}$ on $B$ boundary \cite{cm}, and the variation
  of $B_{ij}$ produces 
\begin{equation}
  \label{spNS}
  \delta B_{ij} = \frac{2}{N} u_{[{i}}\sigma_{{j}]}^{a} \delta(NC_{a})
  - \frac{2}{N} u_{[{i}}\sigma_{{j}]}^{a}D_{ab}\delta V^{b} +
  \sigma_{[{i}}^{a}\sigma_{{j}]}^{b}\delta D_{ab}.
\end{equation}
Putting Eqs. (\ref{ADMdecom}) and (\ref{spNS}) into the ${\cal T}$
boundary term in Eq. (\ref{variation}) leads us to obtain the surface
energy density ${\cal E}$, the surface momentum density ${\cal
  J}^{a}$, the spatial stress ${\cal S}^{ab}$, the surface NS charge
density ${\cal Q}_{\rm NS}^{a}$, the surface NS momentum density
${\cal J}_{\rm NS}^{b}$, and the surface NS current density ${\cal
  I}_{\rm NS}^{ab}$ as
\begin{eqnarray}
  & &{\cal E} \equiv -\frac{\delta S_{\cal T}}{\delta N} =
  \frac{\sqrt{\sigma}}{\pi} (\Phi k - n^{\alpha}
  \nabla_{\alpha} \Phi), \label{def1} \\
  & & {\cal J}^{a} \equiv \frac{\delta S_{\cal T}}{\delta V^{a}} =
  \frac{\sqrt{\sigma}}{\pi} \Phi \sigma^{i}_{a}K_{ij}n^{j}, \label{def2}\\
  & & {\cal S}^{ab} \equiv \frac{\delta S_{\cal T}}{N\delta
  \sigma_{ab}} = \frac{\sqrt{\sigma}}{2\pi} \left[\Phi(k^{ab} -
  \sigma^{ab}k +\sigma^{ab}(n\cdot a)) + \sigma^{ab}n^{\alpha}\nabla_{\alpha}\Phi\right], \label{def3}\\
  & & ({\cal Q_{\rm NS}})^{a}\equiv -\frac{\delta S_{{\cal
  T}}}{\delta (N C_{a})} = \frac{\sqrt{\sigma}}{2\pi}\Phi n_{i}
  {\mathbb E}^{ia}, \label{def4}\\
  & & ({{\cal J}_{\rm NS}})_{a}\equiv \frac{\delta S_{{\cal
  T}}}{\delta {V}^{a}}=({\cal Q}_{\rm NS})^{b}D_{ba}, \label{def5}\\
  & & ({\cal I}_{\rm NS})^{ab} \equiv -\frac{\delta S_{{\cal
  T}}}{N\delta D_{ab}} = \frac{\sqrt{\sigma}}{4\pi} \Phi
  u^{\alpha} H_{\alpha}^{ab}, \label{def6}
\end{eqnarray}
where ${\mathbb
  E}_{ij}=u^{\lambda}h_{i}^{\mu}h_{j}^{\nu}H_{\mu\nu\lambda}$ is the
  ``electric'' piece of the three-form field strength.

On the other hand, equations of motion (\ref{eqnmot}) yields a BTZ black
hole solution, which is given by
\begin{eqnarray}
  \label{btzsol}
  & &(ds)^2_{\rm BTZ} = -N^{2}(r)d^2t + f^{-2}(r) d^2r + r^2 (d\phi +
  N^{\phi}(r)dt)^2, \nonumber\\
  & &\Phi=\Phi(r),~~ B_{\phi t} = B_{\phi t}(r),
\end{eqnarray}
where the lapse function $N^2(r) = (r^2/l^2 - M) $, the shift vector
${(N^{\phi})}^2=J/2r^2$, the dilaton field $\Phi(r) = 1$, and the NS
two-form field potential $B_{\phi t} (r) = r^2/l$. Duality is a symmetry of
string theory, which maps a solution of the low-energy effective
string equations with a translational symmetry to another solution
\cite{hw}. Therefore, this symmetry yields a dual solution of
Eq. (\ref{btzsol})
\begin{eqnarray}
  \label{dualsol}
  & &(ds)_{d}^2 = - N^2(r) d^2t + f^{-2}(r)d^2r + \frac{1}{r^2}(d\phi +
  B_{\phi t} (r) dt)^2, \nonumber \\
  & &\Phi^{d} = r^2 \Phi, ~~B_{\phi t}^{d} = N^{\phi}(r),
\end{eqnarray}
by applying the dual transformation
\begin{eqnarray}
  \label{dualtrans}
  & &g_{xx}^{d} = g_{xx}^{-1}, ~~g_{x\alpha}^{d} = B_{x\alpha}/g_{xx},
  \nonumber \\
  & &g_{\alpha\beta}^{d} = g_{\alpha\beta} - (g_{x\alpha}g_{x\beta} -
  B_{x\alpha}B_{x\beta})/g_{xx},\nonumber \\
  & &B_{x\alpha}^{d} = g_{x\alpha}/g_{xx}, ~~B_{\alpha\beta}^{d} =
  B_{\alpha\beta} - 2g_{x[\alpha}B_{\beta]x}/g_{xx}, \nonumber \\
  & & \Phi^{d} = g_{xx}\Phi,
\end{eqnarray}
where $x$ is a direction of translational symmetry ($\phi$ in our
case) and the superscript $d$ denotes a dual variable. From Eqs. (\ref{btzsol}) and (\ref{dualsol}), the dual
properties of the quasilocal energy density, momentum density, and
NS charge density are obtained by the straightforward calculation of
Eqs. (\ref{def1}), (\ref{def2}), and (\ref{def4}),
\begin{eqnarray}
  \label{dualrel1}
  & &{\cal E} = {\cal E}^{d} = - \frac{f}{\pi}\partial_{r}(r\Phi),
  \nonumber \\
  & &{\cal J}_{\phi} = -({\cal Q}_{\rm NS}^{d})^{\phi} = - \frac{r^3
  f}{2\pi N} \Phi \partial_{r}N^{\phi}, \nonumber \\
  & &({\cal Q}_{\rm NS})^{\phi} = - {\cal J}^{d}_{\phi} =
  \frac{f\Phi}{2\pi Nr}\partial_{r}B_{\phi t}.
\end{eqnarray}
Furthermore, the dual invariance between quasilocal stress density and
dilaton pressure density is satisfied with the combination of both
quantities as
\begin{equation}
  \label{stresspressure}
   {\cal S}^{ab} \delta\sigma_{ab} + \Upsilon \delta\Phi = {\cal
  S}_{d}^{ab} \delta\sigma_{ab}^{d} + \Upsilon_{d}
  \delta\Phi^{d} = \frac{f\partial_{r}N}{2\pi r N} \Phi
  \delta\sigma_{ab} + \frac{f}{\pi}\left(1+\frac{r\partial_{r}N}{N}\right)\delta\Phi,
\end{equation}
where the dilaton pressure density is defined as $\Upsilon \equiv
N^{-1}\Pi_{\rm dil}$.

\subsection{Moving observers and quasilocal quantities}
An extension to the most general case of the quasilocal formalism can
be easily established by assuming that the gravitational system has
non-orthogonal boundaries as shown in FIG. \ref{foliation}. It amounts
to replacing the spatial boundary term ${\cal T}$ in the starting
action (\ref{action}) by 
\begin{equation}
  \label{boostT}
  -\frac{1}{\pi}\int_{\bar{\cal T}}d^2x \sqrt{-\bar{\gamma}} \Phi \bar{\Theta}.
\end{equation}
The variation of this action is written as the similar expression of
Eq. (\ref{variation}) just replaced by ``barred'' expression in $\bar{\cal
  T}$ boundary term, and the boost term $-1/2\pi \int_{B} dx
\sqrt{\sigma} \Phi 2\delta\theta$ is added, where the boost
parameter ${\rm tanh}\theta = v$. The conjugate momenta on $\bar{\cal T}$
boundary are also given as the ``barred'' variables,
\begin{eqnarray}
  \label{barmomT}
  & &\bar{\Pi}^{ij} = \frac{\sqrt{-\bar{\gamma}}}{2\pi}\left[\Phi(\bar{\Theta}^{ij} -
  \bar{\gamma}^{ij}\bar{\Theta}) + \bar{\gamma}^{ij}
  \bar{n}^{\alpha}\nabla_{\alpha}\Phi\right],\nonumber \\
  & &\bar{\Pi}_{\rm dil} =
  \frac{\sqrt{-\bar{\gamma}}}{\pi}(\Phi^{-1}\bar{n}^{\alpha}\nabla_{\alpha}\Phi -
  \bar{\Theta}),\nonumber \\
  & &\bar{\Pi}_{\rm NS}^{ij} = -\frac{\sqrt{-\bar{\gamma}}}{4\pi}\Phi \bar{n}^{\alpha}H^{ij}_{\alpha}.
\end{eqnarray}
The ADM splitting of the induced metric on $\bar{\cal T}$ boundary is
given by the ``barred'' expression of Eq. (\ref{ADMdecom}) while the
induced metric $h_{ij}$ on
$\Sigma$ boundary is splitted by
\begin{equation}
  \label{AdMdecomh}
  \delta h_{ij} = \frac{2}{M} n_{i}n_{j} \delta M + \frac{2}{M}
  \sigma_{a(i}n_{j)} \delta W^{a} + \sigma^{a}_{(i}\sigma^{b}_{j)} \delta\sigma_{ab},
\end{equation}
where $M$ is a lapse function and $W^{a}$ is a shift vector on
$\Sigma$ boundary. Putting these splittings of metrics into the
boundary actions of Eq. (\ref{variation}) yields
\begin{eqnarray}
  \label{btput}
  \int_{\bar{\cal T}} d^2x \bar{\Pi}^{ij} \delta \bar{\gamma}_{ij} &=& -
  \frac{1}{\pi}\int_{\bar{\cal T}} d^2x \sqrt{\sigma} \left[
  \left(\Phi (\gamma k + \gamma v \ell ) -
  \bar{n}^{\alpha}\nabla_{\alpha}\Phi\right) \delta \bar{N} -
  \Phi(\sigma_{a}^{i}K_{ij}n^{j} - \partial_{a}\theta)
  \delta\bar{V}^{a}\nonumber \right. \\& & \left.- \frac{\bar{N}}{2}\left(\Phi\left\{ \gamma
  (k^{ab} - k \sigma^{ab}) + \gamma v (\ell^{ab} - \ell \sigma^{ab}) +
  (\bar{n}\cdot \bar{a})\sigma^{ab}\right\} +
  \bar{n}^{\alpha}\nabla_{\alpha}\Phi \sigma^{ab}\right)\delta\sigma_{ab}\right],
\end{eqnarray}
and
\begin{eqnarray}
  \label{btput2}
  \int_{\Sigma} d^2x P^{ij}\delta h_{ij} &=& \frac{1}{\pi}
  \int_{\Sigma} d^2x \sqrt{\sigma}\left[(\Phi\ell -
  u^{\alpha}\nabla_{\alpha}\Phi)\delta M - \sigma_{a}^{i}\Phi
  K_{ij}n^{j}\delta W^{a} \nonumber \right. \\  &-& \left. \frac{M}{2}\left(\Phi(\ell^{ab} - \ell\sigma^{ab} - (u\cdot b)\sigma^{ab}) +
  u^{\alpha}\nabla_{\alpha}\Phi \sigma^{ab}\right)\delta \sigma_{ab}\right].
\end{eqnarray}
Here we define the quasilocal energy density, the tangential momentum
density, and the spatial stress seen by moving observers in the``barred''
frame as
\begin{eqnarray}
  \label{barredqt}
& & \bar{\cal E} = -\frac{\delta S_{\bar{\cal T}}}{\delta \bar{N}} =
  \frac{\sqrt{\sigma}}{\pi} \left[\Phi \left(\gamma k + \gamma v \ell\right) -
  \bar{n}^{\alpha}\nabla_{\alpha}\Phi\right], \nonumber \\
& & \bar{\cal J}_{a} = \frac{\delta S_{\bar{\cal T}}}{\delta
  \bar{V}^{a}} = \frac{\sqrt{\sigma}}{\pi} \Phi \left(\sigma_{a}^{i}
  K_{ij}n^{j} - \partial_{a} \theta\right), \nonumber \\
& &\bar{\cal S}^{ab} = \frac{\delta S_{\bar{\cal T}}}{\bar{N}\delta
  \sigma_{ab}} = \frac{\sqrt{\sigma}}{2\pi} \left[\Phi\left\{\gamma
  (k^{ab} - k \sigma^{ab}) + \gamma v (\ell^{ab} - \ell \sigma^{ab}) +
  (\bar{n} \cdot\bar{a})\sigma^{ab}\right\} +
  \sigma^{ab}\bar{n}^{\alpha}\nabla_{\alpha}\Phi \right],
\end{eqnarray}
and the quasilocal normal momentum density, the tangential momentum
density, and the temporal stress seen by static observers in the
``unbarred'' frame as
\begin{eqnarray}
  \label{unbarredqt}
  & & {\cal J}_{\vdash} = -\frac{\delta S_{\Sigma}}{\delta M} = -
  \frac{\sqrt{\sigma}}{\pi}\left[ \Phi \ell -
  u^{\alpha}\nabla_{\alpha}\Phi\right], \nonumber \\
  & & {\cal J}_{a} = -\frac{\delta S_{\Sigma}}{\delta W^{a}} =
  \frac{\sqrt{\sigma}}{\pi}\Phi
  \sigma_{a}^{i}K_{ij}n^{j},\nonumber \\
  & & {\Delta}^{ab} = \frac{\delta S_{\Sigma}}{M\delta \sigma^{ab}} =
  -\frac{\sqrt{\sigma}}{2\pi}\left[\Phi\left\{\ell^{ab} - \ell
  \sigma^{ab} - (u\cdot b)\sigma^{ab}\right\} + \sigma^{ab}u^{\alpha}\nabla_{\alpha}\Phi\right].
\end{eqnarray}
In addition, the dilaton pressure scalar densities on $\bar{\cal T}$ and $\Sigma$
boundaries are calculated as
\begin{eqnarray}
  \label{scalarden}
  & &\bar{\Upsilon} = \bar{N}^{-1}\bar{\Pi}_{\rm dil} =
  \frac{\sqrt{\sigma}}{\pi}\left(\Phi^{-1}\bar{n}^{\alpha}\nabla_{\alpha}\Phi
  - \gamma k - \gamma v\ell + (\bar{n}\cdot\bar{a})\right),\nonumber \\
  & &{\cal Z} = M^{-1} P_{\rm dil} = -
  \frac{\sqrt{\sigma}}{\pi}\left(\Phi^{-1}u^{\alpha}\nabla_{\alpha}\Phi
  - \ell - (u\cdot b)\right).
\end{eqnarray}

As for the NS charge part, we have the variation of action for a NS three-form
field strength $H_{\mu\nu\rho}$,
\begin{equation}
  \label{3form}
  \delta S_{\rm NS} = \int_{\cal M} d^3x \sqrt{-g}\left(\Xi_{\rm
  NS}\right)^{\mu\nu}\delta B_{\mu\nu} + \int_{\Sigma}d^2x P_{\rm
  NS}^{ij}\delta B_{ij} + \int_{\bar{\cal T}} d^2x \bar{\Pi}_{\rm
  NS}^{ij} \delta B_{ij},
\end{equation}
where the equation of motion $(\Xi_{\rm
  NS})^{\mu\nu}$ and the canonical momenta $P_{\rm
  NS}^{ij}$ and $\bar{\Pi}_{\rm
  NS}^{ij}$ on both boundaries are given by Eqs. (\ref{eqnmot}),
  (\ref{momS}), and (\ref{barmomT}), respectively. Note that the three-form
  field strength $H_{\mu\nu\rho}$ is usually decomposed into
  ``electric'' and ``magnetic'' components on a spacelike
  hypersurface, ${\mathbb{E}}_{ij} =
  h_{i}^{\mu}h_{j}^{\nu}H_{\mu\nu\rho}u^{\rho}$ and
  ${\mathbb{B}}=-\epsilon^{\mu\nu\rho\lambda}H_{\mu\nu\rho}u_{\lambda}/6$,
  respectively, and it can be shown that $H^{2} = 6{\mathbb{B}}^2 -
  3{\mathbb E}_{ij}{\mathbb E}^{ij}$. As shown in Eq. (\ref{spNS}), the
  two-form field potential $B_{ij}$ can be decomposed on the $\Sigma$
  boundary into
  \begin{equation}
    \label{Bdecom}
    \delta B_{ij} = - \frac{2}{M} n_{[i}\sigma_{j]}^{a}\delta(ME_{a}) -
    \frac{2}{M} n_{[i}\sigma_{j]}^{a}D_{ab}\delta W^{b} +
    \sigma_{[i}^{a}\sigma_{j]}^{b} \delta D_{ab},
  \end{equation}
where $B_{ij} = -2n_{[i}\sigma_{j]}^{a} E_{a} +
\sigma_{[i}^{a}{\sigma_{j]}}^{b}D_{ab}$ and $E_{a} = \sigma_{a}^{i}B_{ij}n^{j}$, and the field decomposition
on $\bar{\cal T}$ boundary is written as a similar form of
Eq. (\ref{spNS})
\begin{equation}
  \label{barspNS}
  \delta B_{ij} = \frac{2}{\bar{N}} \bar{u}_{[{i}}\sigma_{{j}]}^{a}
  \delta(\bar{N}\bar{C}_{a})  - \frac{2}{\bar{N}} \bar{u}_{[{i}}\sigma_{{j}]}^{a}D_{ab}\delta \bar{V}^{b} +
  \sigma_{[{i}}^{a}\sigma_{{j}]}^{b}\delta D_{ab},
\end{equation}
where $\bar{C}_{ab} = \sigma_{a}^{i} B_{ij} \bar{u}^{j}$. 
Hereafter, substituting Eqs. (\ref{Bdecom}) and (\ref{barspNS}) into
Eq. (\ref{3form}) and imposing the equations of motion gives
\begin{eqnarray}
  \label{NSsub}
  \delta S_{\rm NS} &=& \int_{\Sigma} d^2x \left[ - ({\cal J}_{\rm
  NS})_{a}\delta W^{a} - ({\cal Q}_{\rm NS})^{a}\delta(ME_{a}) + M
  ({\cal I}_{\rm NS}^{\rm u})^{ab}\delta D_{ab}\right] \nonumber \\
&+&  \int_{\bar{\cal T}} d^2x \left[ (\bar{{\cal J}}_{\rm
  NS})_{a}\delta \bar{V}^{a} - (\bar{{\cal Q}}_{\rm NS})^{a}\delta(\bar{N}\bar{C}_{a}) - \bar{N}
  (\bar{{\cal I}}_{\rm NS}^{\rm n})^{ab}\delta D_{ab}\right],
\end{eqnarray}
where the surface NS charge density, the surface NS momentum density,
and the surface NS current density seen by static(``unbarred'') and
moving(``barred'') observers are
\begin{equation}
  \label{NSquantunbar}
  ({\cal Q}_{\rm NS})^{a} = \frac{\sqrt{\sigma}}{2\pi} \Phi n_{i}
  {\mathbb E}^{ia}, ~({\cal J}_{\rm NS})_{a} = ({\cal Q}_{\rm NS})^{b}D_{ba},~ ({\cal I}_{\rm NS}^{\rm u})^{ab} = \frac{\sqrt{\sigma}}{4\pi} \Phi
  u^{\alpha} H_{\alpha}^{ab},
\end{equation}
and
\begin{equation}
  \label{NSquantbar}
  (\bar{{\cal Q}}_{\rm NS})^{a} = \frac{\sqrt{\sigma}}{2\pi} \Phi \bar{n}_{i}
  \bar{{\mathbb E}}^{ia}, ~(\bar{{\cal J}}_{\rm NS})_{a} = (\bar{{\cal Q}}_{\rm NS})^{b}D_{ba},~ (\bar{{\cal I}}_{\rm NS}^{\rm n})^{ab} = \frac{\sqrt{\sigma}}{4\pi} \Phi
  \bar{n}^{\alpha} H_{\alpha}^{ab},
\end{equation}
respectively. Note that the surface NS charge density and the surface
NS momentum density in the boosted and unboosted frames are obtained
from the each boundary term, but the surface NS current densities are
divided by two terms projected with respect to the unit normal vectors
$u^{\mu}$ and $\bar{n}^{\mu}$ in Eqs. (\ref{NSquantunbar}) and
(\ref{NSquantbar}). The notations of quasilocal quantities used in
this paper in the boosted and unboosted frames are summarized in TABLE
\ref{table2}. 

\begin{table}
\caption{Notations for quasilocal quantities in boosted and unboosted frames}
\begin{tabular}{lccccccccc}\label{table2}
  field contents & \vbox{\hbox{$\bar{N}$}\hbox{$N$}} & \vbox{\hbox{$\bar{M}$}\hbox{$M$}} & $\bar{V}^{a}$ & $W^{a}$ & $\sigma_{ab}$ & $ME_{a}$ &
  $\bar{N}\bar{C}_{a}$ & $D_{ab}$ & $\Phi$  \\ \tableline
\vbox{\hbox{quantities}\hbox{in ``barred'' frame}} &
  $\bar{\cal E}$ & $\bar{\cal J}_{\vdash}$ & \vbox{\hbox{$\bar{\cal J}_{a}$,}\hbox{$(\bar{{\cal J}}_{\rm NS})_{a}$}} & &
 \vbox{\hbox{$\bar{{\cal S}}^{ab}$,}\hbox{$\bar{\Delta}^{ab}$}} & & $(\bar{{\cal Q}}_{\rm NS})^{a}$
  & \vbox{\hbox{$(\bar{{\cal I}}_{\rm NS}^{\rm n})^{ab}$,}\hbox{$(\bar{{\cal I}}_{\rm NS}^{\rm
  u})^{ab}$}} & $\bar{\Upsilon}$ \\
\vbox{\hbox{quantities}\hbox{in ``unbarred'' frame}} &
  ${\cal E}$ & ${\cal J}_{\vdash}$ &  & \vbox{\hbox{${\cal J}_{a}$,}\hbox{$({\cal J}_{\rm NS})_{a}$}} &
  \vbox{\hbox{${\cal S}^{ab}$,}\hbox{${\Delta}^{ab}$}} & $({\cal Q}_{\rm NS})^{a}$
  &  & \vbox{\hbox{$({\cal I}_{\rm NS}^{\rm n})^{ab}$,}\hbox{$({\cal I}_{\rm NS}^{\rm
  u})^{ab}$}} & ${\cal Z}$ \\
\end{tabular}
\end{table}

\subsection{Boost relations and duality of quasilocal variables}
The quasilocal quantities seen by moving observers are connected by
those seen by static observers through the boost relations. We have
quasilocal quantities in ``unbarred'' frame as follows
\begin{eqnarray}
  \label{unbarred}
  & &{\cal E} = \frac{\sqrt{\sigma}}{\pi} \left[\Phi k -
  n^{\alpha}\nabla_{\alpha}\Phi\right],\nonumber \\
  & &{\cal J}_{\vdash} =- \frac{\sqrt{\sigma}}{\pi}\left[\Phi\ell -
  u^{\alpha}\nabla_{\alpha}\Phi\right], \nonumber \\
  & &{\cal J}_{a} = \frac{\sqrt{\sigma}}{\pi}\Phi \sigma_{a}^{i} K_{ij}n^{j}, \nonumber \\
  & &{\cal S}^{ab} =
  \frac{\sqrt{\sigma}}{2\pi}\left[\Phi\left(k^{ab} - \sigma^{ab}k +
  (n\cdot  a)\sigma^{ab}\right) + \sigma^{ab}n^{\alpha}\nabla_{\alpha}\Phi \right], \nonumber \\
  & &{\Delta}^{ab} =
  -\frac{\sqrt{\sigma}}{2\pi}\left[\Phi\left(\ell^{ab} -
  \ell\sigma^{ab} - (u\cdot b)\sigma^{ab}\right) +
  \sigma^{ab}u^{\alpha}\nabla_{\alpha}\Phi\right],
\end{eqnarray}
and these are simply converted into the quasilocal quantities seen in
``barred'' frame as
\begin{eqnarray}
  \label{barred}
  & &\frac{\pi}{\sqrt{\sigma}}\bar{{\cal E}} = \Phi \bar{k} -
  \bar{n}^{\alpha}\nabla_{\alpha}\Phi = \Phi(\gamma k + \gamma v
  \ell)-\bar{n}^{\alpha}\nabla_{\alpha}\Phi,\nonumber \\
  & &\frac{\pi}{\sqrt{\sigma}}\bar{{\cal J}_{\vdash}} =
  -\Phi\bar{\ell} + \bar{u}^{\alpha}\nabla_{\alpha}\Phi =
  -\gamma(\Phi\ell - u^{\alpha}\nabla_{\alpha}\Phi) - \gamma v (\Phi k
  - n^{\alpha}\nabla_{\alpha}\Phi ), \nonumber \\
  & &\frac{\pi}{\sqrt{\sigma}}\bar{{\cal J}_{a}} = \Phi \sigma_{a}^{i}
  \bar{K}_{ij} \bar{n}^{j} = \Phi(\sigma_{a}^{i} K_{ij} n^{j} -
  \partial_{a} \theta),
\end{eqnarray}
using the relations of unit normal vectors in Eq. (\ref{relunit}). The
spatial and temporal stress tensors are given as
\begin{eqnarray}
  \label{sptempstress}
  \frac{2\pi}{\sqrt{\sigma}} \bar{\cal S}^{ab} &=& \Phi
  (\bar{k}^{ab} - \sigma^{ab}\bar{k}
  +(\bar{n}\cdot\bar{a})\sigma^{ab}) +
  \sigma^{ab}\bar{n}^{\alpha}\nabla_{\alpha}\Phi \nonumber \\
  &=& \gamma\left[\Phi\left\{k^{ab}-\sigma^{ab}k +
  (n\cdot a)\sigma^{ab}\right\} +
  \sigma^{ab} n^{\alpha}\nabla_{\alpha}\Phi\right] \nonumber \\ &+& \gamma v
  \left[\Phi\left\{\ell^{ab} - \ell\sigma^{ab} - (u\cdot
  b)\sigma^{ab}\right\} +
  \sigma^{ab}u^{\alpha}\nabla_{\alpha}\Phi\right] +
  \Phi(\bar{u}\cdot\nabla\theta)\sigma^{ab}, \nonumber \\
  \frac{2\pi}{\sqrt{\sigma}}\bar{\Delta}^{ab} &=&
  -\Phi(\bar{\ell}^{ab} - \bar{\ell}\sigma^{ab} - (\bar{u}\cdot
  \bar{b})\sigma^{ab}) -
  \sigma^{ab}\bar{u}^{\alpha}\nabla_{\alpha}\Phi \nonumber \\
  &=& \gamma\left[ - \Phi\left\{\ell^{ab} -
  \ell\sigma^{ab} - (u\cdot b)\sigma^{ab} \right\} - \sigma^{ab}
  u^{\alpha}\nabla_{\alpha}\Phi \right]\nonumber \\ &+& \gamma v \left[ -
  \Phi\left\{ k^{ab} - k \sigma^{ab} + (n\cdot a)\sigma^{ab} \right\}
  - \sigma^{ab}n^{\alpha}\nabla_{\alpha}\Phi \right] + \Phi(\bar{n}\cdot\nabla\theta)\sigma^{ab},
\end{eqnarray}
by using Eq. (\ref{relunit}), and
$(\bar{n}\cdot\bar{a}) = \gamma(n\cdot a) - \gamma v (u\cdot b) +
\bar{u}\cdot \nabla\theta$ and $(\bar{u}\cdot\bar{b}) = \gamma(u\cdot
b) - \gamma v (n\cdot a) - \bar{n}\cdot \nabla\theta$.
Therefore, the boost relations between the surface energy
density, the tangential momentum density, the normal momentum density,
the spatial stress, and the temporal stress in the boosted and
unboosted frames are obtained as
\begin{eqnarray}
  \label{boostrel}
  & &\bar{{\cal E}} = \gamma {\cal E} - \gamma v {\cal J}_{\vdash},
  \nonumber \\
  & &\bar{{\cal J}_{\vdash}} = \gamma {\cal J}_{\vdash} - \gamma v
  {\cal E}, \nonumber \\
  & &\bar{{\cal J}_{a}} = {\cal J}_{a} - \frac{1}{\pi}\Phi
  \partial_{a}\theta,\nonumber \\
  & &\bar{{\cal S}}^{ab} = \gamma {\cal S}^{ab} - \gamma v
  {\Delta}^{ab} + \frac{\sqrt{\sigma}}{2\pi}\Phi(\bar{u}\cdot \nabla \theta)\sigma^{ab}, \nonumber \\
  & &\bar{{\Delta}}^{ab} = \gamma {\Delta}^{ab} - \gamma v {\cal
  S}^{ab} + \frac{\sqrt{\sigma}}{2\pi} \Phi(\bar{n}\cdot \nabla \theta)\sigma^{ab},
\end{eqnarray}
by using Eqs. (\ref{unbarred}), (\ref{barred}), and
(\ref{sptempstress}), and the boost relations for the quasilocal NS charge
densities, NS momentum densities, and NS current densities are simply
given as 
\begin{eqnarray}
  \label{NSboostrel}
& &(\bar{{\cal Q}}_{\rm NS})^{a} = \gamma^{2} ({\cal Q}_{\rm NS})^{a} +
  2\gamma^{2}v^{2} n_{\mu}({\cal I}_{\rm NS}^{\rm u})^{\mu a},
  \nonumber \\
& &(\bar{{\cal J}}_{\rm NS})_{b} = \gamma^{2} ({\cal J}_{\rm
  NS})_{b} + 2 \gamma^{2} v^{2} n_{\mu}({\cal I}_{\rm NS}^{\rm
  u})^{\mu a} D_{ab}, \nonumber \\
& &(\bar{{\cal I}}_{\rm NS}^{\rm n})^{ab} = \gamma({\cal I}_{\rm
  NS}^{\rm n})^{ab} + \gamma v ({\cal I}_{\rm NS}^{\rm u})^{ab},
\end{eqnarray}
by means of Eqs. (\ref{NSquantunbar}) and (\ref{NSquantbar}). Note that the boost invariance of the tangential momentum density in
Eq. (\ref{boostrel}) and NS charge density in
Eq. (\ref{NSboostrel}) are straightforwardly calculated for the metric
(\ref{btzsol}) as $(\bar{{\cal Q}}_{\rm NS})^{\phi} = ({\cal Q}_{\rm
  NS})^{\phi}$ and $\bar{{\cal J}}_{\phi} = {\cal J}_{\phi}$,
respectively, which are expected results since the
only motion in our case is perpendicular to the angular direction.   

Let us now show the duality relations
between the surface energy densities $\bar{\cal E}$ and $\bar{{\cal E}}^{d}$,
the tangential momentum densities $\bar{{\cal J}}_{\vdash}$ and $\bar{{\cal J}}_{\vdash}^{d}$,
the normal momentum densities $\bar{{\cal J}}_{\phi}$ and $\bar{{\cal
  J}}_{\phi}^{d}$, and the NS charge densities $(\bar{{\cal Q}}_{\rm
NS})^{\phi}$ and $(\bar{{\cal Q}}_{\rm NS}^{d})^{\phi}$. 
Using the boost relations in Eqs. (\ref{boostrel}) and
(\ref{NSboostrel}), the dual relations are given as
\begin{eqnarray}
  \label{duality}
  & & \bar{{\cal E}} = \bar{{\cal E}^{d}}, \nonumber \\
  & & \bar{{\cal J}}_{\vdash} = \bar{{\cal J}}_{\vdash}^{d}, \nonumber
  \\
  & & \bar{{\cal J}}_{\phi} = - (\bar{{\cal Q}}_{\rm NS}^{d})^{\phi},
  \nonumber \\
  & & (\bar{{\cal Q}}_{\rm NS})^{\phi} = - \bar{{\cal J}}_{\phi}^{d},
\end{eqnarray}
and note that these relations are exactly same forms with those of the
orthogonal boundary case. Notice that Eq. (\ref{duality}) shows that
the dual properties between the quasilocal variables are still valid
regardless of observers who measure the quasilocal variables in their
own frames. 

Next let us focus on the dual invariance of the spatial and temporal
stress densities and dilaton pressure densities. Basically, the
quantity $(n\cdot a)$ has a dual invariance for the
metrics (\ref{btzsol}) and (\ref{dualsol}), and it yields
$(\bar{n}\cdot\bar{a}) = \gamma(n\cdot a) =
(\bar{n}\cdot\bar{a})_{d}$. In the ``barred'' frame, the combination
of spatial stress and dilaton pressure density satisfies the dual
invariance, which is given by
\begin{eqnarray}
  \label{dilspatial}
  \bar{\cal S}^{ab} \delta \sigma_{ab} + \bar{\Upsilon} \delta
  \Phi &=& \gamma ({\cal S}^{ab} \delta \sigma_{ab} + {\Upsilon} \delta
  \Phi) + \frac{\gamma v}{2\pi} \partial_{r}\theta
  \left(\frac{f}{r}\Phi \delta\sigma_{ab} + 2rf \delta\Phi\right)
  \nonumber \\
  &=& \gamma ({\cal S}^{ab}_{d}\delta\sigma_{ab}^{d} + \Upsilon_{d}
  \delta \Phi^{d}) + \frac{\gamma v}{2\pi}\partial_{r}\theta \left(r^3 f\Phi
  \delta\sigma_{ab}^{d} + \frac{2f}{r}\delta\Phi^{d}\right) \nonumber
  \\
  &=& \bar{\cal S}^{ab}_{d} \delta \sigma_{ab}^{d} + \bar{\Upsilon}_{d} \delta
  \Phi^{d},
\end{eqnarray}
and the additional dual relation for the temporal stress density
$\Delta^{ab}$ and the dilaton scalar density ${\cal Z}$ is obtained by
a simple calculation,
\begin{equation}
  \label{additional}
  \Delta^{ab} \delta\sigma_{ab} + {\cal Z}\delta\Phi = \Delta^{ab}_{d}
  \delta\sigma_{ab}^{d} + {\cal Z}_{d}\delta\Phi^{d} = 0.
\end{equation}

As a result, the whole quasilocal quantities are reformulated by the
double-foliation of quasilocal analysis with non-orthogonal
boundaries, and the relevant boost relations are
presented. Furthermore, the dual properties for quasilocal variables
are still valid even in the moving observer's frame. 

\section{Discussions \hfil{}}\label{sec:IV}
We have studied the duality of quasilocal energy and charges for the
(2+1)-dimensional dilatonic gravitational system with
non-orthogonal boundaries by use of the double-foliation of spacetime
manifold ${\cal M}$. The quasilocal variables including the surface
energy density, momentum densities, spatial and temporal stresses, and
the quantities related to the NS three-form field strength have been
presented and the dual relations between those quantities have been proposed.
In this approach, the boosting is confined
to the radial direction, so the angular momentum densities and NS
charge density are independent of the boost factor $\gamma$ while the
energy density is mixed with the tangential momentum
density ${\cal J}_{\vdash}$. In other words, those quantities are
naturally expected to have a general covariance under Lorentz-type
transformations. 

On the other hand, for a non-compact spacetime, quasilocal quantities
are not well defined in the limit that a finite boundary $R$
goes to infinity. This unexpected inconsistency can be removed by
introducing reference background spacetimes with an action $S_{0}$ and
the physical action can be defined as $S_{\rm phys.} = S -S_{0}$. However,
this reference background spacetime action does not guarantee to preserve the
covariance of quasilocal quantities since those variables in the
reference background spacetimes will transform with a different
velocity comparing with the velocity of the quasilocal surface in the
given spacetimes. Nevertheless, the reference background spacetimes action does
not alter the dual properties of quasilocal quantities.
In fact, the action (\ref{action}) is reduced to the effective action $S_{eff} = 1/2\pi \int_{\cal M} d^3x
  \sqrt{-g}\left(R+2/l^2\right)$ by imposing solution of the
  dilaton field and NS three-form field strength \cite{kal}. It evidently
  describes an AdS$_3$ spacetimes and the gravitational counter term
  for AdS$_3$ spacetimes can be considered as a reference background
spacetime action. For
an AdS spacetime, the counter term action can be constructed by algorithmic
procedure and it is uniquely determined \cite{kls}. The counter term
action of Eq. (\ref{action}) is written as 
$S_{ct} = -1/{\pi l} \int_{\bar{\cal T}} d^2x
\sqrt{-\bar{\gamma}}\Phi$, where $\Phi(r) = 1$ for the BTZ black hole,
which is compatible with the action shown in Refs. \cite{kls,bk}, and
it is invariant under the dual transformation
(\ref{dualtrans}). More precisely, the reference background action
gives the reference energy density, the reference spatial stress, and
the reference dilaton pressure density as $\bar{{\cal E}}_{0} =
\sqrt{\sigma}\Phi / \pi l$, $\bar{{\cal S}}^{ab}_{0} = -
\sqrt{\sigma}\sigma^{ab}\Phi/ 2\pi l$, and $\bar{\Upsilon}_{0} = -
\sqrt{\sigma}/\pi l$, respectively. A short glance of ${\cal E}_{0}$
shows that it is invariant under boosting and dual transformation, i.e., $\bar{{\cal E}}_{0}={\cal
  E}_{0}$ and $\bar{{\cal E}}_{0} = \bar{{\cal E}}_{0}^{d} $. In addition, the
combination of $\bar{{\cal S}}^{ab}_{0} \delta \sigma_{ab} +
\bar{\Upsilon}_{0}\delta\Phi$ is also invariant under the dual
transformation (\ref{dualtrans}). Therefore, the physical quasilocal quantities
by subtracting the values of the reference background spacetimes
inevitably satisfies the usual properties of dual transformations
to any observers whether they are moving or not.
\vspace{2cm}
\acknowledgments{
J. J. Oh would like to thank to J. Ho and G. Kang for useful
discussions and K. H. Yee is grateful to P. P. Jung for helpful
comments. This work was supported by the grant No. 2000-2-11100-002-5
from the Basic Research Program of the Korea Science and Engineering
Foundation.}

\end{document}